\documentclass[sigconf,authorversion,nonacm]{acmart}

\acmConference[MSR 2024]{21st International Conference on Mining Software Repositories}{April 2024}{Lisbon, Portugal}
\AtBeginDocument{%
  \providecommand\BibTeX{{%
    \normalfont B\kern-0.5em{\scshape i\kern-0.25em b}\kern-0.8em\TeX}}}

\usepackage{xspace}
\usepackage{tikz,pifont,pgfplots}
\usepackage{xcolor, colortbl}
\usepackage{makecell,multirow,diagbox}
\usepackage[many]{tcolorbox}
\usepackage{graphicx}%
\usepackage{adjustbox}
\usepackage{amsmath,amsfonts}%
\usepackage{mathrsfs}%
\usepackage[title]{appendix}%
\usepackage{xcolor}%
\usepackage{textcomp}%
\usepackage{manyfoot}%
\usepackage{booktabs}%
\usepackage{algorithm}%
\usepackage{algorithmicx}%
\usepackage{algpseudocode}%
\usepackage{listings}%
\usetikzlibrary{positioning}
\usepackage{arydshln}
\usepackage{url}
\usepackage{multirow}
\usepackage{subcaption}
\usepackage{pdflscape}
\usepackage{tabularx}

\pgfplotsset{compat=1.18}

\newcommand*{\ie}{i.e.,\@\xspace}
\newcommand*{\eg}{e.g.,\@\xspace}

\newcommand*{\tool}{PlayMyData\@\xspace}
\newcommand*{\ig}{IGDB\@\xspace}
\newcommand*{\hl}{HLTB\@\xspace}
\newcommand*{\yt}{YouTube\@\xspace}
\newcommand*{\pl}{PlayStation\@\xspace}
\newcommand*{\xb}{Xbox\@\xspace}
\newcommand*{\nt}{Nintendo\@\xspace}

\lstdefinestyle{searchstringstyle}{
  basicstyle=\ttfamily\scriptsize,
breaklines=true,                 
  captionpos=b,                    
  numbers=none,                    
  numbersep=5pt,                  
  showspaces=false,                
  showstringspaces=false,
  showtabs=false,                  
  tabsize=2,
  frame=single
}


\definecolor{verylightgray}{gray}{0.92}
\definecolor{ao(english)}{rgb}{0.0, 0.5, 0.0}

\definecolor{deepblue}{rgb}{0,0,0.5}
\definecolor{deepred}{rgb}{0.6,0,0}
\definecolor{deepgreen}{rgb}{0,0.5,0}
\definecolor{shadecolor}{gray}{0.9}

\colorlet{shadecolor}{verylightgray}
\colorlet{framecolor}{black}

    \lstdefinestyle{searchstringstyle}{
	basicstyle=\ttfamily\scriptsize,
	breakatwhitespace=false,         
breaklines=true,                 
	captionpos=b,                 
	keepspaces=true,                 
	numbers=none,                    
	numbersep=4pt,                  
	showspaces=false,                
	showstringspaces=false,
	showtabs=false,                  
	tabsize=2,
	frame=single
}

\newtcolorbox{shadedbox}{
drop shadow southeast,
enhanced jigsaw,
colback=white,
boxrule=0.80pt,
left=0.3em,
right=0.3em,
top=0.1em,
bottom=0.05em
}
\newcommand*{\numberGames}{99,864\@\xspace}
\newcommand*{\numberVideos}{43,812\@\xspace}
\newcommand*{\numberScreen}{443,630\@\xspace}

\definecolor{mygreen}{rgb}{0,0.6,0}
\definecolor{mygray}{rgb}{0.95,0.95,0.95}
\definecolor{myred}{rgb}{0.5,0,0}

\definecolor{verylightgray}{gray}{0.92}
\definecolor{ao(english)}{rgb}{0.0, 0.5, 0.0}

\colorlet{punct}{red!60!black}
\definecolor{background}{HTML}{EEEEEE}
\definecolor{delim}{RGB}{20,105,176}
\colorlet{numb}{magenta!60!black}

\lstdefinelanguage{json}{
	basicstyle=\footnotesize\ttfamily,
	numbers=left,
	numberstyle=\scriptsize,
	stepnumber=1,
	numbersep=8pt,
	showstringspaces=false,
	breaklines=true,
	frame=lines,
	backgroundcolor=\color{background},
	literate=
	*{0}{{{\color{numb}0}}}{1}
	{1}{{{\color{numb}1}}}{1}
	{2}{{{\color{numb}2}}}{1}
	{3}{{{\color{numb}3}}}{1}
	{4}{{{\color{numb}4}}}{1}
	{5}{{{\color{numb}5}}}{1}
	{6}{{{\color{numb}6}}}{1}
	{7}{{{\color{numb}7}}}{1}
	{8}{{{\color{numb}8}}}{1}
	{9}{{{\color{numb}9}}}{1}
	{:}{{{\color{punct}{:}}}}{1}
	{,}{{{\color{punct}{,}}}}{1}
	{\{}{{{\color{delim}{\{}}}}{1}
	{\}}{{{\color{delim}{\}}}}}{1}
	{[}{{{\color{delim}{[}}}}{1}
	{]}{{{\color{delim}{]}}}}{1},
}

\begin{document}


\title{\tool: Enabling mining tasks on multi-platform videogames}

\title{\tool: a curated multi-purpose dataset of videogames}

\title{\tool: a curated dataset of multi-platform video games}
\author{Andrea D'Angelo}
\email{andrea.dangelo6@graduate.univaq.it}
\orcid{0000-0002-0577-2494}
\affiliation{%
  \institution{Università degli studi dell'Aquila}
\streetaddress{Via Vetoio 2, 67100}
  \city{L'Aquila}
  \country{Italy}
}

\author{Claudio Di Sipio}
\email{claudio.disipio@univaq.it}
\orcid{0000-0001-9872-9542}
\affiliation{%
  \institution{Università degli studi dell'Aquila}
\streetaddress{Via Vetoio 2, 67100}
  \city{L'Aquila}
  \country{Italy}
}

\author{Cristiano Politowski}
\email{cristiano.politowski@umontreal.ca}
\affiliation{%
  \institution{DIRO, University of Montreal}
  \city{Montreal}
  \country{Canada}
}

\author{Riccardo Rubei}
\email{riccardo.rubei@univaq.it}
\orcid{0000-0001-9622-5949}
\affiliation{%
  \institution{Università degli studi dell'Aquila}
\streetaddress{Via Vetoio 2, 67100}
  \city{L'Aquila}
  \country{Italy}
}








\renewcommand{\shortauthors}{D'Angelo et al.}

\begin{abstract}
Being predominant in digital entertainment for decades, video games have been recognized as valuable software artifacts by the software engineering (SE) community just recently. Such an acknowledgment has unveiled several research opportunities, spanning from empirical studies to the application of AI techniques for classification tasks. In this respect, several curated game datasets have been disclosed for research purposes even though the collected data are insufficient to support the application of advanced models or to enable interdisciplinary studies. Moreover, the majority of those are limited to PC games, thus excluding notorious gaming platforms, \eg \pl, \xb, and \nt. 
In this paper, we propose \tool, a curated dataset composed of \numberGames multi-platform games gathered by \ig website. By exploiting a dedicated API, we collect relevant metadata for each game, \eg description, genre, rating, gameplay video URLs, and screenshots. Furthermore, we enrich \tool with the timing needed to complete each game by mining the \hl website. To the best of our knowledge, this is the most comprehensive dataset in the domain that can be used to support different automated tasks in SE. More importantly, \tool can be used to foster cross-domain investigations built on top of the provided multimedia data. 

\end{abstract} 






\maketitle

\section{Introduction}
\label{sec:introduction}
Over the last decades, digital entertainment has become a billion-dollar business and it is still growing \cite{gibson2022relationship}. In this respect, the videogame industry plays a crucial role by offering tons of new products every year. This expanding market segment has attracted the academic interest in different domains, \eg behavior analysis \cite{granic2014benefits,quwaider2019impact,kuhn2019does}, gender bias detection\cite{forni2020horizon,madden2021you} and addiction \cite{griffiths2009videogame,wood2008problems}. Just recently, the software engineering (SE) community has started to consider games as knowledgeable software artifacts \cite{pascarella_how_2018}, thus opening several research opportunities, \eg code smell detection \cite{10.1145/3563214}, sentiment analysis \cite{thompson2017sentiment,ardianto2020sentiment}, and serious games \cite{10.1145/3573074.3573096}. 
Similar to traditional software artifacts, the main challenge remains the data gathering since most of the well-known digital stores and websites \cite{steam,playstore,metacritic} collect unstructured data or don't support mining activities with a proper API, thus requiring extra development efforts.  


%
To fill this gap, we propose \tool, a curated videogame dataset of \numberGames games belonging to main gaming platforms (called \textit{platforms} hereon) \ie \pl, \xb, \nt, and PC, stored on \ig website \footnote{\url{https://www.igdb.com/}}. The rationale behind this choice is \textit{i)} it offers a dedicated API \cite{igdb-api_getting_2023} to collect the needed data and \textit{ii)} the mined data are reusable for non-commercial usage \footnote{\url{https://www.twitch.tv/p/it-it/legal/developer-agreement/}}, thus fostering the reproducibility of the results. In the scope of the paper, we mined all relevant metadata to support automated text classification, \eg descriptions, game genres, and ratings. In addition, we collect \numberVideos video gameplay URLs and \numberScreen screenshots that can be used to support computer vision tasks. We further enhance \tool with game completion times, \ie the number of hours to complete a game,  stored on \hl website\footnote{\url{howlongtobeat.com}} by using a community-based API \cite{michele_howlongtobeat_2023}. To this end, we query \hl by applying the Levensthein distance \cite{levenshtein} to the game title, with the aim of reducing possible false positives. To the best of our knowledge, \tool is the first multi-platform dataset that includes completion times. Even though the primary usage of \tool is towards the support of automated approaches, we can envision different potential usages in the social domain by means of the collected multimedia data, \ie screenshots and gameplay videos. The mined dataset is made publicly available at \url{https://zenodo.org/records/10262075}.



\section{\tool collection}
\label{sec:proposed}
\begin{figure}
    \centering
    \includegraphics[width=1\linewidth]{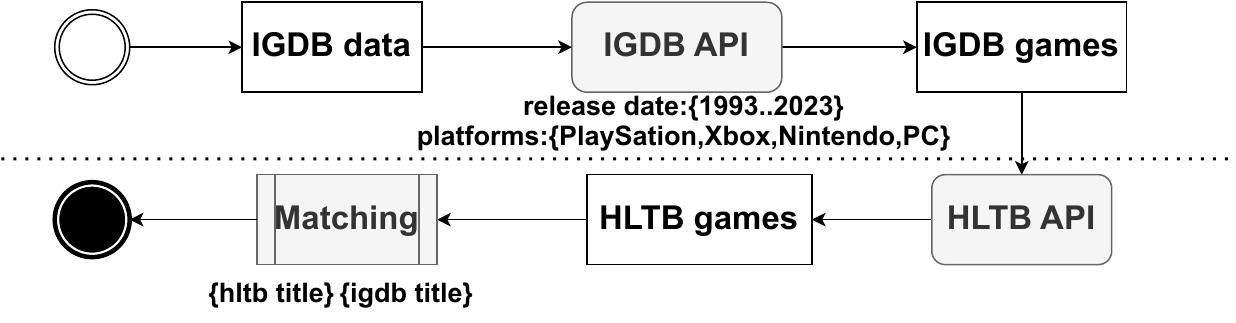}
    \caption{The \tool collection process.}
    \label{fig:dataMining}
\end{figure}

Figure \ref{fig:dataMining} depicts the data collection process of the two selected websites, \ie \ig and \hl. 
Concerning the former, we collect relevant metadata to our purpose on \ig including the screenshots and the URLs gameplay videos. Afterwards, we searched the titles of the retrieved games on the \hl platform and we matched them using the Levenstein similarity function. 

\subsection{\tool data model} \label{sec:data_model}

\begin{figure}
    \centering
    \includegraphics[width=1\linewidth]{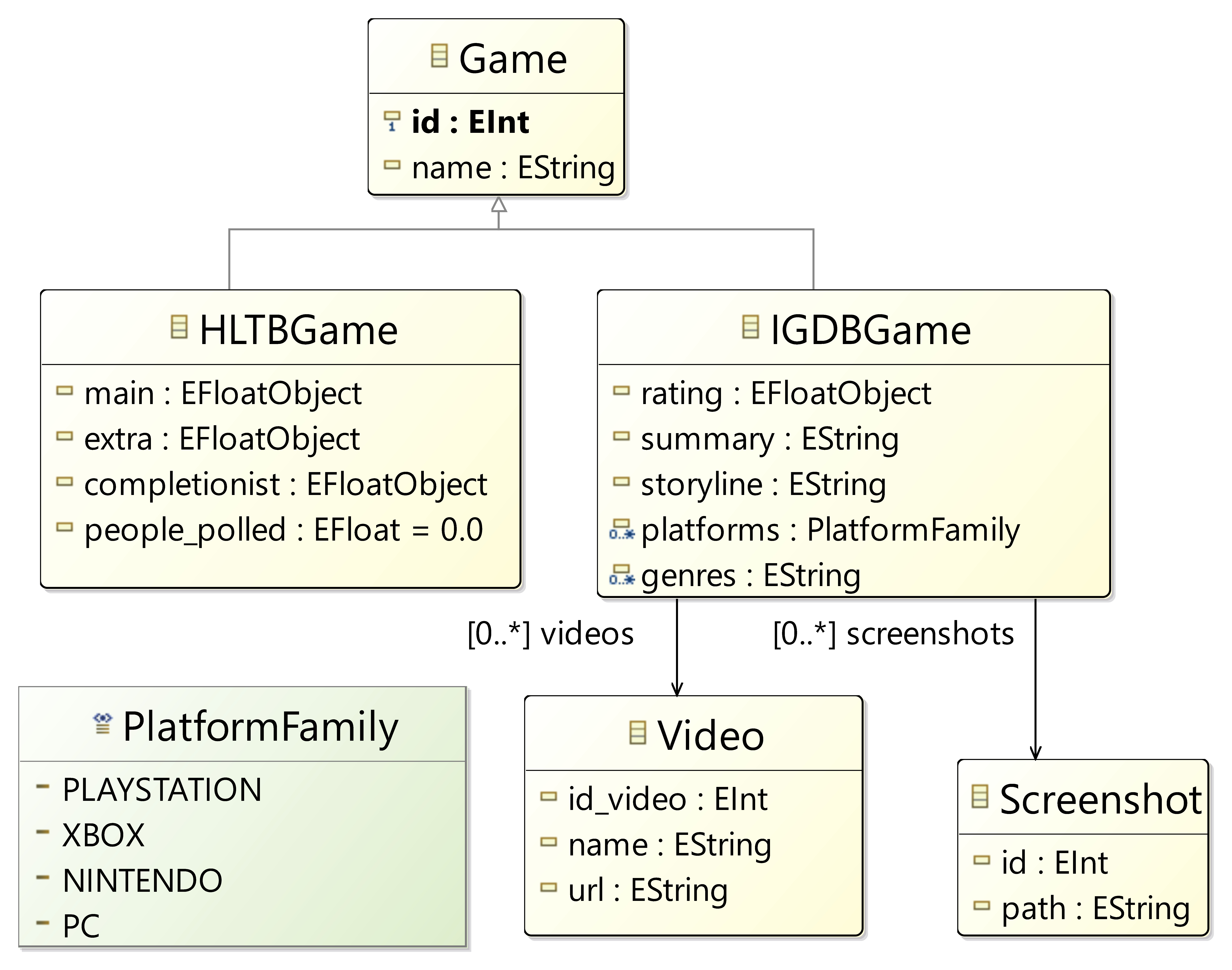}
    \caption{\tool data model.}
    \label{fig:dataModel}
\end{figure}

Figure \ref{fig:dataModel} depicts the \tool data model. The main concept is the \texttt{Game} class that represents a game, with a unique integer id and a name, \ie the title of the game. Since we mined games from two different storage,  the \texttt{IGDBGame} and \texttt{HLTBGame} classes represent the collected data from \ig and \hl respectively. An \texttt{IGDBGame} contains a \texttt{summary} of the game and a more succinct description, \ie the \texttt{storyline}. Furthermore, the average rating assigned by the users is represented by the attribute \texttt{rating}. A game can belong to  23 different \texttt{genres} available on \ig\footnote{The reader can refer to the complete list of genres in the provided archive}. Similarly, the same game has been released on different \texttt{platforms}. We define the \texttt{PlatformFamily} enumeration that describes the considered platforms \ie \pl, \xb, \nt, and PC. Roughly speaking,  we consider all the platform versions for a specific family. For instance, \pl literal indicates all the \pl version, \eg PSone, PS2. Each \texttt{IGDBGame} can possibly have a set of gameplay video and screenshots represented by the \texttt{Video} and \texttt{Screenshot} classes.

Similarly, the \texttt{HLTBGame} class represents specific information concerning the gameplay time expressed in hours for each game. In particular, \hl stores the time needed to complete the \texttt{main} story, \texttt{extra} that includes the side-quests times, and the time to collect all the game achievements, \ie \texttt{completionist} attribute. In addition, we collect the number of users that submit their completion time, \ie \texttt{people\_polled} attribute. In particular, this attribute measures the popularity and engagement levels of each game within the gaming community.

\subsection{IGDB mining} \label{sec:igbd}

To collect data from \ig, we rely on the dedicated service \cite{igdb-api_getting_2023} that facilitates the data-gathering phase through an authorized API.
Mining all the games stored on \ig is out of the scope of this paper. Thus, we focus on games released from 1$^{st}$ January 1993 up to 30$^{th}$ November 2023 belonging to the best-selling platforms\footnote{\url{https://www.techradar.com/news/best-consoles}}, \ie \pl, \xb, \nt, and PC. To this end, we exploit the \textit{first release} field available on \ig. 
For each game, we fetch the metadata represented in the data model (cf. Section \ref{sec:data_model}) using the query shown in Listing \ref{lst:queryGame}:

\begin{lstlisting}[label=lst:queryGame,caption= The IGDB query to collect metadata.,style=searchstringstyle]
"fields [list fields]; where platforms= [platform ids] and first_release_date >= min date and  first_release_date < max date; sort rating desc; limit 500;"
\end{lstlisting}
 
The \textit{list fields} and \textit{platforms ids} are the attributes of an \texttt{IGDBGame} as described in Section \ref{sec:data_model} and the list of unique IDs for each platform respectively. Meanwhile, \textit{min date} and \textit{max date} represent the time interval for each considered year. 
To handle the API rate limit, we set this interval equal to one month to fetch the maximum number of games for each month, by setting the \textit{limit} equal to 500. The retrieved games are eventually ordered by the average user ratings, from the most rated to the least one.  

In addition, we downloaded the screenshots and the gameplay video metadata for each game using two different sets of queries available here \cite{riccardorubeimsr2024-data-showcase} that are similar to the one shown in Listing \ref{lst:queryGame}. We downloaded the screenshot files for all games since are in the \textit{thumbnails} size, \ie the entire size of the dataset is still manageable. Contrariwise, we opt to fetch only the \yt URLs of the gameplay videos. Even though \tool does not include the corresponding \texttt{.MP4} files, we provide a dedicated functionality that makes use of the PyTube library \footnote{\url{https://pytube.io/en/latest/}}. We made available an explanatory video in the supporting GitHub repository \cite{riccardorubeimsr2024-data-showcase}.

For each video, we store the name as appears on \ig. This information can be further exploited to categorize the videos into trailers and actual gameplay released by the software house. 

Overall, we collected \numberGames games, \numberScreen screenshots, and \numberVideos video URLs. Metadata related to games and videos are stored in CSV format while we group the screenshots using the genre of the corresponding game. 

\subsection{HLTB mining} \label{sec:hltb}

The second phase involves the data collection from \hl. In the scope of this work, we customize the community-built API \cite{michele_howlongtobeat_2023} to retrieve the data related to the game completion as discussed in Section \ref{sec:data_model}.

The conceived components execute a POST request on the \hl website using the title of the game as shown in Listing \ref{lst:queryhltb}.
\\
\begin{lstlisting}[label=lst:queryhltb,caption= The explanatory HLTB query.,style=searchstringstyle]
    {"searchType": "games", "searchTerms": ["game_title"], "searchPage": 1, "size": 20}
\end{lstlisting}
 In particular, the \textit{game\_title} parameter is the name attribute of an \ig game as described in Figure \ref{fig:dataModel} while the \textit{search page} is set to 1 since we limit ourselves to the first result page. The \textit{size} parameter represents the number of possible matches for that tile. In the scope of our analysis, we select the first one from this list. 
 This component eventually retrieves a set of possible candidates that need to be mapped to \ig games collected in the previous phase. 

\subsection{Matching \ig and \hl games} \label{sec:matching}

To match the collected data gathered from \hl, we exploit the common element between the two platforms, \ie the game title. However, this information may not perfectly match those stored on \hl or may not be present at all, resulting in the retrieval of an entirely different game.

To overcome these issues, when merging the data from the two different sources, we compute the Levenshtein distance \cite{levenshtein}, a widely-used technique to measure the edit distance between two strings according to the standard formula: 
\smallskip

    \begin{equation}\label{eqn:Levenshtein}
    L_{s_1, s_2}(i, j) = 
        \begin{cases}
        \max(i, j) & \text{if } \min(i, j) = 0,\\
        L_{s_1, s_2}(i-1, j-1) & \text{if } s_1[i] = s_2[j],\\
        \min
            \begin{cases}
            L_{s_1, s_2}(i - 1, j) + 1\\
            L_{s_1, s_2}(i, j - 1) + 1\\
            L_{s_1, s_2}(i - 1, j - 1) + 1 & \end{cases} & \text{otherwise.}        
        \end{cases}
    \end{equation}
\smallskip



where $s_1$ and $s_2$ are the \ig and \hl game titles respectively, and i and j are their indices, starting from their length. 
In the scope of our work, we empirically set the similarity threshold equal to 3. On one hand, a higher threshold might lead to the inclusion of titles that are too different, potentially merging unrelated games. On the other hand, a lower threshold could be overly restrictive, excluding valid matches where titles vary slightly between the two databases. 
Roughly speaking, a Levenshtein distance of 3 allows for minor typographical differences or variations in game titles, thus maintaining an adequate level of accuracy during the matching phase. In total, we merged completion times data for 35,815 different games from \hl.

\section{\tool overview}
\label{sec:overview}

This section presents an overview of the collected data by providing basic descriptive statistics. In addition, we analyze the collected games over time by considering their completion time.

\subsubsection*{Descriptive statistic:}

\tool contains \numberGames unique games over on a total number of 266,072 currently stored on \ig\footnote{\url{https://www.igdb.com/about}}. By carefully inspecting the gathered data, we counted missing values for the \textit{storyline}, \textit{genres}, and \textit{ratings} metadata. In particular, the storyline description is missing for only 445 entries while a relevant number of games have no rating, \ie 32,859 games. A possible explanation is that only popular games are often rated by the users. This is confirmed by the \hl since the \textit{people\_polled} value is equal to 0 for 62,450 games, meaning that any \hl user didn't submit its time completion for these games. 

\begin{figure}[htbp]
    \centering
    \begin{subfigure}[b]{0.5\linewidth}
        \includegraphics[width=1\linewidth]{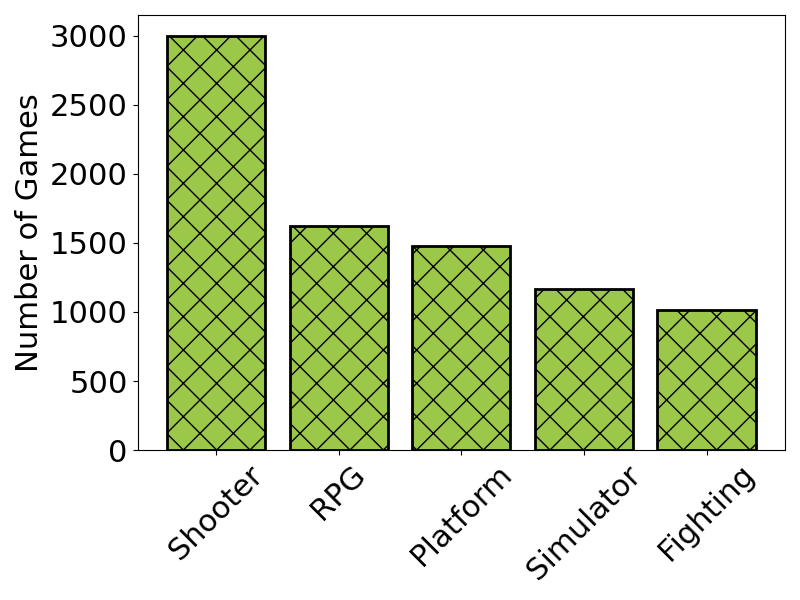}
        \caption{Xbox}
        \label{fig:xbox}
    \end{subfigure}%
    \begin{subfigure}[b]{0.5\linewidth}
        \includegraphics[width=\linewidth]{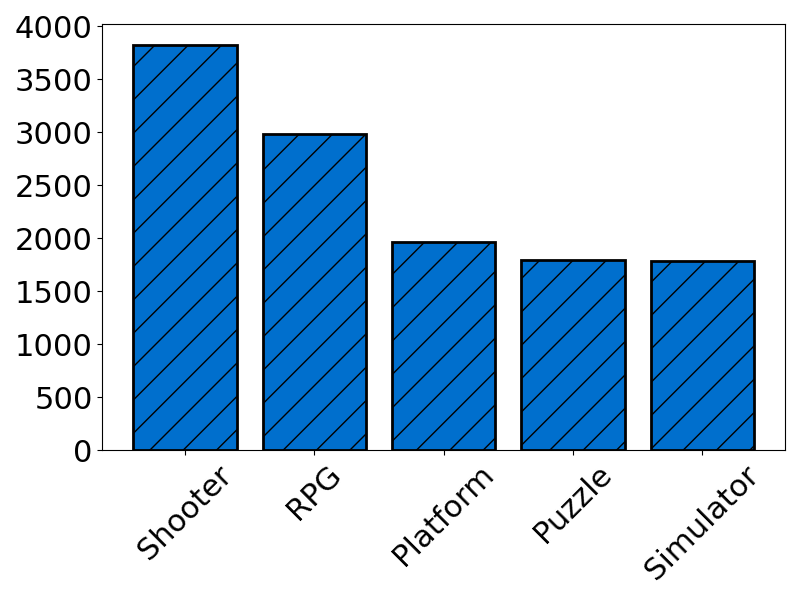}
        \caption{Playstation}
        \label{fig:playstation}
    \end{subfigure}

    \begin{subfigure}[b]{0.5\linewidth}
        \includegraphics[width=\linewidth]{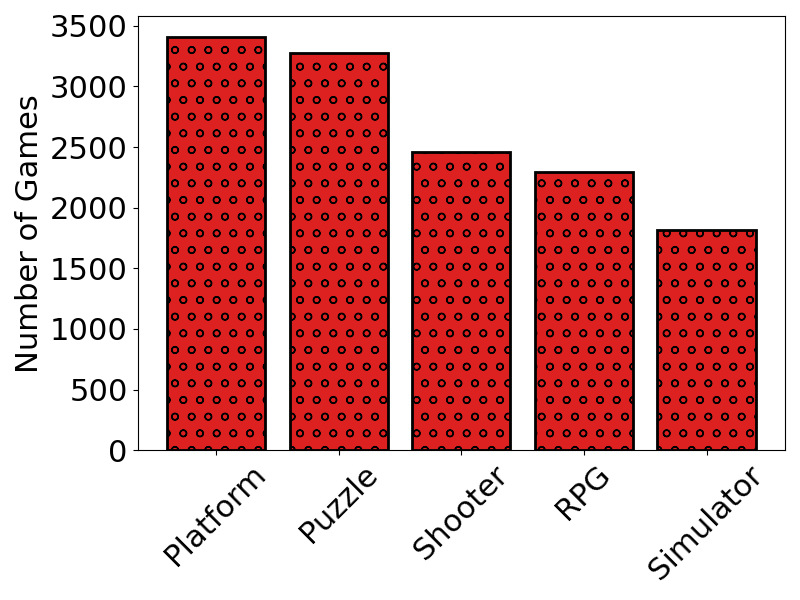}
        \caption{Nintendo}
        \label{fig:nintendo}
    \end{subfigure}%
    \begin{subfigure}[b]{0.5\linewidth}
        \includegraphics[width=\linewidth]{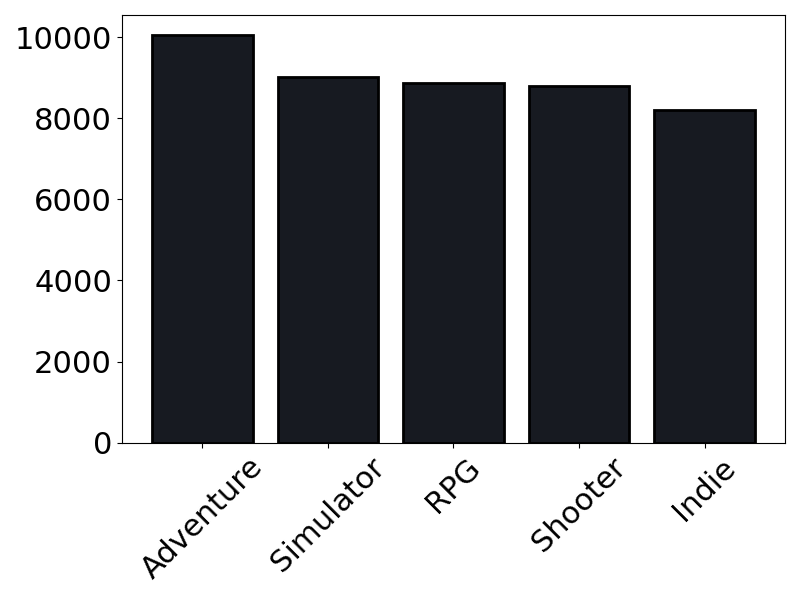}
        \caption{PC}
        \label{fig:pc}
    \end{subfigure}
    \caption{Most spread genres grouped by platform family.}
    \label{fig:games_platforms}
\end{figure}

We further analyzed the data checking the top five genres for each game considered platform as shown in Figure \ref{fig:games_platforms}. Games tagged as \textit{Shooter} are more common on \xb and \pl as depicted in Figure \ref{fig:xbox} and Figure \ref{fig:playstation}. On the other hand, \nt games are mostly in the \textit{Platform} genre, followed by \textit{Puzzle} games. \textit{Indie} games only appear on PC, with more than 8,200 titles, thus confirming the recent trend that the releases are ``heavily skewed by small indie projects'' \footnote{\url{https://tinyurl.com/2bw7zw4t}}. Overall, the most popular genres are \textit{Adventure}, \textit{Shooter}, \textit{RPG}, and \textit{Simulator}, with more than 10K games each.

\subsubsection*{Game completion over time:} 

\begin{figure}[!htb]
    \centering
    \includegraphics[width=1\linewidth]{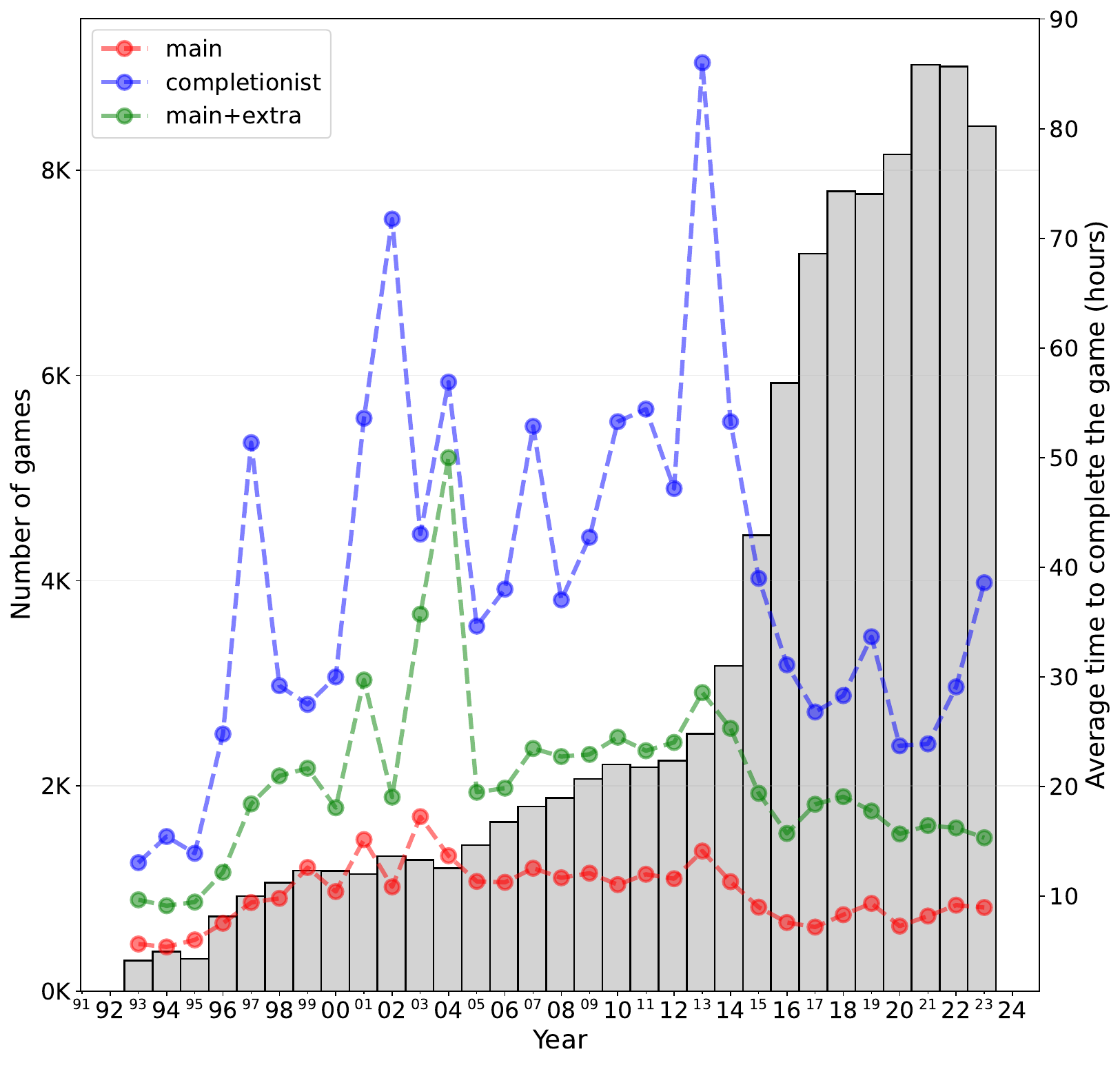}
    \caption{Evolution of game completion over time.}
    \label{fig:hist-games-year}
\end{figure}

We explored the \hl data and checked how the completion time has evolved over the years. \autoref{fig:hist-games-year} depicts the number of games released per year and the corresponding completion time considering the three different attributes, \ie main, extra, and completionist described in Section \ref{sec:data_model}. As expected, the number of released games per year has grown during the considered time window, \ie from less than 1,000 games released in 1993 to more than 8,000 in 2022 on average. 

Even though the completion time follows a similar trend on average, the number of hours collected from \hl doesn't grow at the same scale. The conducted analysis shows that completing a game nowadays requires less time compared to the middle 2000s, but still more than in the early 1990s. Even though recent findings show that the games are becoming longer "to beat" \footnote{\url{https://tinyurl.com/4xxk7dty}}, this trend doesn't affect all the genres. In addition, the analysis conducted on our dataset reveals a peak in 2013, in particular for collecting all the game achievements, \ie the \texttt{completionist} attribute reached 80 hours on average. Investigating the possible reasons behind such a phenomenon is beyond the scope of this paper and is left as a possible future work. 



\section{Research opportunities}
\label{sec:opportunities}

\noindent
\textbf{Text classification:} The existing literature exploits textual data to automatically categorize games \cite{DBLP:journals/corr/abs-2105-05674,rubei2021auryga,bertens_machine-learning_2018} or provide a personalized list of games \cite{pathak_generating_2017,liu_recommenderplus_2022}.
Compared to existing approaches, \tool collects a large set of multi-platform games and the corresponding relevant data to facilitate the application of advanced techniques, \eg pre-trained models. Furthermore, we augment the \ig data with the completion time for each game collected from the \hl storage, thus enabling the possibility of including such data to enhance automatic classification or developing recommender systems for games.  

\noindent
\textbf{Image classification tasks:} In the realm of image classification tasks, games remain an underexplored domain despite the widespread popularity of such models. Notably, in \cite{jiang_deep_2020}, images were leveraged to feed a multi-modal model designed for genre classification in video games. 
\tool provides a vast repository of thumbnail screenshots, all standardized to the same size of 90x90 pixels. This consistency in image dimensions adheres to best practices established in the field \cite{alexnet,vggnet}. The screenshots provided by \tool serve as a helpful resource to build models able to classify a game by genre or other attributes from still images.



\noindent
\textbf{Analyzing gameplay videos:} Gameplay and video contribution have been employed to detect errors or anomalies \cite{guglielmi2023using} and for classification purposes \cite{8463434}. Recently, gameplay videos have been employed to derive user behavior in gamification \cite{hadi_mogavi_your_2023}. Even though \tool collects only the YouTube URL videos, we offer an easy way to download and increase the collection of those artifacts.

\section{Threats to validity}
\label{sec:threats}
In this section, we briefly discuss the limitations of our
paper. Concerning the \textit{internal validity}, one possible threat is the selected time window,\ie we consider only games released from 1993 to 2023.  
To mitigate any bias possible, we considered the most spread platforms,\ie \pl, \xb, \nt, and PC, thus building a representative enough dataset composed of 37\% of the games stored on \ig. In addition, we provide only the gameplay video URLs instead of the actual .MP4 file. To handle this, we provide a dedicated function to download the video from \yt, thus allowing interested researchers to get all the stored videos. 
Threats to \textit{external validity} are related to missing game information of \tool, \ie videos, ratings, and time completion. In particular, we may miss relevant data from \hl since the stored games are different from \ig. We opt for the Levensthein distance to match the game titles and minimize the number of missing entries. 



\section{Related works}
\label{sec:relatedWorks}
\citet{jiang_deep_2020} proposed a multi-modal classifier based on \ig games. The authors used the cover image textual data to classify 50,000 PC games using their genres. \citet{politowski2020dataset} manually analyzed postmortems of released games and created a dataset with 200 postmortems from 1998 to 2018, extracting 1,035 SE-related problems with traceability links. \citet{melhart2022arousal} provided a dataset for affect modeling named AGAIN. The dataset consists of 37 hours of video footage from 1,116 gameplay sessions played by 124 participants. ViGGO \cite{juraska2019viggo} is a corpus for data-to-text Natural Language Generation that includes 100 games and about 7000 pairs of structured meaning representations. SIMS4ACTION \cite{roitberg2021let} dataset collects more than 10 hours of gameplay of The Sims 4 game to simulate Activities of Daily Living (ADL). 

Compared to the abovementioned approaches, \tool collects a larger number of games, \ie \numberGames, considering additional gaming platforms apart from PC, \ie \pl, \xb, and \nt. Furthermore, we consider the completion time gathered from \hl, screenshots, and gameplay video URLs.

\section{Conclusion and future works}
\label{sec:conclusion}
Intending to support the intersection between SE and entertainment, we present \tool, a well-structured dataset composed of multi-platform games belonging to the main gaming platforms, \ie \xb, \pl, \nt, and PC. First, we collected \numberGames games that date back to 30 years ago available on the \ig website including their relevant metadata, \numberScreen screenshots, and \numberVideos URLs of gameplay videos. 
Second, completion time has been collected from \hl, a dedicated website that discloses such data, by applying the Levensthein distance on titles to reduce any possible bias. Although some data are missing, \tool can be used to support cross-domain investigations, moving from social science to the application of automated techniques. 
In future works, we plan to fill in the missing entries and further expand the offered metadata, \eg including game reviews. In addition, the older platforms can be included in the analysis, \eg Sega or Amiga, to investigate the evolution of games over time. Last but not least, we can exploit multimedia data, \ie screenshots and videos, to feed multimodal approaches to understand their impact on the classification task.

\clearpage
\bibliographystyle{ACM-Reference-Format}
\bibliography{main}


\begin{thebibliography}{35}


\ifx \showCODEN    \undefined \def \showCODEN     #1{\unskip}     \fi
\ifx \showDOI      \undefined \def \showDOI       #1{#1}\fi
\ifx \showISBNx    \undefined \def \showISBNx     #1{\unskip}     \fi
\ifx \showISBNxiii \undefined \def \showISBNxiii  #1{\unskip}     \fi
\ifx \showISSN     \undefined \def \showISSN      #1{\unskip}     \fi
\ifx \showLCCN     \undefined \def \showLCCN      #1{\unskip}     \fi
\ifx \shownote     \undefined \def \shownote      #1{#1}          \fi
\ifx \showarticletitle \undefined \def \showarticletitle #1{#1}   \fi
\ifx \showURL      \undefined \def \showURL       {\relax}        \fi
\providecommand\bibfield[2]{#2}
\providecommand\bibinfo[2]{#2}
\providecommand\natexlab[1]{#1}
\providecommand\showeprint[2][]{arXiv:#2}

\bibitem[met(2023)]%
        {metacritic}
 \bibinfo{year}{2023}\natexlab{}.
\newblock \bibinfo{title}{Video {Game} {Reviews}, {Articles}, {Trailers} and more}.
\newblock
\newblock
\urldef\tempurl%
\url{https://www.metacritic.com/game/}
\showURL{%
\tempurl}
\newblock
\shownote{[last accessed on 2023-12-06]}.


\bibitem[Ardianto et~al\mbox{.}(2020)]%
        {ardianto2020sentiment}
\bibfield{author}{\bibinfo{person}{Rian Ardianto}, \bibinfo{person}{Tri Rivanie}, \bibinfo{person}{Yuris Alkhalifi}, \bibinfo{person}{Fitra~Septia Nugraha}, {and} \bibinfo{person}{Windu Gata}.} \bibinfo{year}{2020}\natexlab{}.
\newblock \showarticletitle{Sentiment analysis on E-sports for education curriculum using naive Bayes and support vector machine}.
\newblock \bibinfo{journal}{\emph{Jurnal Ilmu Komputer dan Informasi}} \bibinfo{volume}{13}, \bibinfo{number}{2} (\bibinfo{year}{2020}), \bibinfo{pages}{109--122}.
\newblock


\bibitem[Bertens et~al\mbox{.}(2018)]%
        {bertens_machine-learning_2018}
\bibfield{author}{\bibinfo{person}{Paul Bertens}, \bibinfo{person}{Anna Guitart}, \bibinfo{person}{Pei~Pei Chen}, {and} \bibinfo{person}{Africa Perianez}.} \bibinfo{year}{2018}\natexlab{}.
\newblock \showarticletitle{A machine-learning item recommendation system for video games}. In \bibinfo{booktitle}{\emph{2018 IEEE Conference on Computational Intelligence and Games (CIG)}}. IEEE, \bibinfo{pages}{1--4}.
\newblock


\bibitem[Bucchiarone et~al\mbox{.}(2023)]%
        {10.1145/3573074.3573096}
\bibfield{author}{\bibinfo{person}{Antonio Bucchiarone}, \bibinfo{person}{Kendra M.~L. Cooper}, \bibinfo{person}{Dayi Lin}, \bibinfo{person}{Edward~F. Melcer}, {and} \bibinfo{person}{Kelvin Sung}.} \bibinfo{year}{2023}\natexlab{}.
\newblock \showarticletitle{Games and Software Engineering: Engineering Fun, Inspiration, and Motivation}.
\newblock \bibinfo{journal}{\emph{SIGSOFT Softw. Eng. Notes}} \bibinfo{volume}{48}, \bibinfo{number}{1} (\bibinfo{date}{jan} \bibinfo{year}{2023}), \bibinfo{pages}{85–89}.
\newblock
\showISSN{0163-5948}
\urldef\tempurl%
\url{https://doi.org/10.1145/3573074.3573096}
\showDOI{\tempurl}


\bibitem[Forni(2020)]%
        {forni2020horizon}
\bibfield{author}{\bibinfo{person}{Dalila Forni}.} \bibinfo{year}{2020}\natexlab{}.
\newblock \showarticletitle{Horizon Zero Dawn: The educational influence of video games in counteracting gender stereotypes}.
\newblock \bibinfo{journal}{\emph{Transactions of the Digital Games Research Association}} \bibinfo{volume}{5}, \bibinfo{number}{1} (\bibinfo{year}{2020}).
\newblock


\bibitem[Gibson et~al\mbox{.}(2022)]%
        {gibson2022relationship}
\bibfield{author}{\bibinfo{person}{Erin Gibson}, \bibinfo{person}{Mark~D Griffiths}, \bibinfo{person}{Filipa Calado}, {and} \bibinfo{person}{Andrew Harris}.} \bibinfo{year}{2022}\natexlab{}.
\newblock \showarticletitle{The relationship between videogame micro-transactions and problem gaming and gambling: A systematic review}.
\newblock \bibinfo{journal}{\emph{Computers in Human Behavior}}  \bibinfo{volume}{131} (\bibinfo{year}{2022}), \bibinfo{pages}{107219}.
\newblock


\bibitem[Granic et~al\mbox{.}(2014)]%
        {granic2014benefits}
\bibfield{author}{\bibinfo{person}{Isabela Granic}, \bibinfo{person}{Adam Lobel}, {and} \bibinfo{person}{Rutger~CME Engels}.} \bibinfo{year}{2014}\natexlab{}.
\newblock \showarticletitle{The benefits of playing video games.}
\newblock \bibinfo{journal}{\emph{American psychologist}} \bibinfo{volume}{69}, \bibinfo{number}{1} (\bibinfo{year}{2014}), \bibinfo{pages}{66}.
\newblock


\bibitem[Griffiths and Meredith(2009)]%
        {griffiths2009videogame}
\bibfield{author}{\bibinfo{person}{Mark~D Griffiths} {and} \bibinfo{person}{Alex Meredith}.} \bibinfo{year}{2009}\natexlab{}.
\newblock \showarticletitle{Videogame addiction and its treatment}.
\newblock \bibinfo{journal}{\emph{Journal of Contemporary Psychotherapy}}  \bibinfo{volume}{39} (\bibinfo{year}{2009}), \bibinfo{pages}{247--253}.
\newblock


\bibitem[Guglielmi et~al\mbox{.}(2023)]%
        {guglielmi2023using}
\bibfield{author}{\bibinfo{person}{Emanuela Guglielmi}, \bibinfo{person}{Simone Scalabrino}, \bibinfo{person}{Gabriele Bavota}, {and} \bibinfo{person}{Rocco Oliveto}.} \bibinfo{year}{2023}\natexlab{}.
\newblock \showarticletitle{Using gameplay videos for detecting issues in video games}.
\newblock \bibinfo{journal}{\emph{Empirical Software Engineering}} \bibinfo{volume}{28}, \bibinfo{number}{6} (\bibinfo{year}{2023}), \bibinfo{pages}{136}.
\newblock


\bibitem[Hadi~Mogavi et~al\mbox{.}(2023)]%
        {hadi_mogavi_your_2023}
\bibfield{author}{\bibinfo{person}{Reza Hadi~Mogavi}, \bibinfo{person}{Chao Deng}, \bibinfo{person}{Jennifer Hoffman}, \bibinfo{person}{Ehsan-Ul Haq}, \bibinfo{person}{Sujit Gujar}, \bibinfo{person}{Antonio Bucchiarone}, {and} \bibinfo{person}{Pan Hui}.} \bibinfo{year}{2023}\natexlab{}.
\newblock \showarticletitle{Your {Favorite} {Gameplay} {Speaks} {Volumes} {About} {You}: {Predicting} {User} {Behavior} and {Hexad} {Type}}. In \bibinfo{booktitle}{\emph{{HCI} in {Games}}} \emph{(\bibinfo{series}{Lecture {Notes} in {Computer} {Science}})}, \bibfield{editor}{\bibinfo{person}{Xiaowen Fang}} (Ed.). \bibinfo{publisher}{Springer Nature Switzerland}, \bibinfo{address}{Cham}, \bibinfo{pages}{210--228}.
\newblock
\showISBNx{978-3-031-35979-8}
\urldef\tempurl%
\url{https://doi.org/10.1007/978-3-031-35979-8_17}
\showDOI{\tempurl}


\bibitem[Horppu et~al\mbox{.}(2021)]%
        {DBLP:journals/corr/abs-2105-05674}
\bibfield{author}{\bibinfo{person}{Ismo Horppu}, \bibinfo{person}{Antti Nikander}, \bibinfo{person}{Elif Buyukcan}, \bibinfo{person}{Jere M{\"{a}}kiniemi}, \bibinfo{person}{Amin Sorkhei}, {and} \bibinfo{person}{Frederick Ayala{-}G{\'{o}}mez}.} \bibinfo{year}{2021}\natexlab{}.
\newblock \showarticletitle{Automatic Classification of Games using Support Vector Machine}.
\newblock \bibinfo{journal}{\emph{CoRR}}  \bibinfo{volume}{abs/2105.05674} (\bibinfo{year}{2021}).
\newblock
\showeprint[arxiv]{2105.05674}
\urldef\tempurl%
\url{https://arxiv.org/abs/2105.05674}
\showURL{%
\tempurl}


\bibitem[igdb API(2023)]%
        {igdb-api_getting_2023}
\bibfield{author}{\bibinfo{person}{igdb API}.} \bibinfo{year}{2023}\natexlab{}.
\newblock \bibinfo{title}{Getting {Started} – {IGDB} {API} docs}.
\newblock
\newblock
\urldef\tempurl%
\url{https://api-docs.igdb.com/#getting-started}
\showURL{%
\tempurl}


\bibitem[Jiang and Zheng(2020)]%
        {jiang_deep_2020}
\bibfield{author}{\bibinfo{person}{Yuhang Jiang} {and} \bibinfo{person}{Lukun Zheng}.} \bibinfo{year}{2020}\natexlab{}.
\newblock \showarticletitle{Deep learning for video game genre classification}.
\newblock \bibinfo{journal}{\emph{arXiv:2011.12143 [cs]}} (\bibinfo{date}{Nov.} \bibinfo{year}{2020}).
\newblock
\urldef\tempurl%
\url{http://arxiv.org/abs/2011.12143}
\showURL{%
\tempurl}
\newblock
\shownote{00001 arXiv: 2011.12143}.


\bibitem[Juraska et~al\mbox{.}(2019)]%
        {juraska2019viggo}
\bibfield{author}{\bibinfo{person}{Juraj Juraska}, \bibinfo{person}{Kevin~K Bowden}, {and} \bibinfo{person}{Marilyn Walker}.} \bibinfo{year}{2019}\natexlab{}.
\newblock \showarticletitle{ViGGO: A video game corpus for data-to-text generation in open-domain conversation}.
\newblock \bibinfo{journal}{\emph{arXiv preprint arXiv:1910.12129}} (\bibinfo{year}{2019}).
\newblock


\bibitem[Krizhevsky et~al\mbox{.}(2017)]%
        {alexnet}
\bibfield{author}{\bibinfo{person}{Alex Krizhevsky}, \bibinfo{person}{Ilya Sutskever}, {and} \bibinfo{person}{Geoffrey~E. Hinton}.} \bibinfo{year}{2017}\natexlab{}.
\newblock \showarticletitle{ImageNet Classification with Deep Convolutional Neural Networks}.
\newblock \bibinfo{journal}{\emph{Commun. ACM}} \bibinfo{volume}{60}, \bibinfo{number}{6} (\bibinfo{date}{may} \bibinfo{year}{2017}), \bibinfo{pages}{84–90}.
\newblock
\showISSN{0001-0782}
\urldef\tempurl%
\url{https://doi.org/10.1145/3065386}
\showDOI{\tempurl}


\bibitem[K{\"u}hn et~al\mbox{.}(2019)]%
        {kuhn2019does}
\bibfield{author}{\bibinfo{person}{Simone K{\"u}hn}, \bibinfo{person}{Dimitrij~Tycho Kugler}, \bibinfo{person}{Katharina Schmalen}, \bibinfo{person}{Markus Weichenberger}, \bibinfo{person}{Charlotte Witt}, {and} \bibinfo{person}{J{\"u}rgen Gallinat}.} \bibinfo{year}{2019}\natexlab{}.
\newblock \showarticletitle{Does playing violent video games cause aggression? A longitudinal intervention study}.
\newblock \bibinfo{journal}{\emph{Molecular psychiatry}} \bibinfo{volume}{24}, \bibinfo{number}{8} (\bibinfo{year}{2019}), \bibinfo{pages}{1220--1234}.
\newblock


\bibitem[Liu(2022)]%
        {liu_recommenderplus_2022}
\bibfield{author}{\bibinfo{person}{Tianrui Liu}.} \bibinfo{year}{2022}\natexlab{}.
\newblock \showarticletitle{{RecommenderPlus}: {New} {Content}-based {User}-centered {Game} {Recommendation} {System}}. In \bibinfo{booktitle}{\emph{2022 3rd {International} {Conference} on {Computer} {Vision}, {Image} and {Deep} {Learning} \& {International} {Conference} on {Computer} {Engineering} and {Applications} ({CVIDL} \& {ICCEA})}}. \bibinfo{pages}{767--770}.
\newblock
\urldef\tempurl%
\url{https://doi.org/10.1109/CVIDLICCEA56201.2022.9825356}
\showDOI{\tempurl}


\bibitem[Madden et~al\mbox{.}(2021)]%
        {madden2021you}
\bibfield{author}{\bibinfo{person}{Daniel Madden}, \bibinfo{person}{Yuxuan Liu}, \bibinfo{person}{Haowei Yu}, \bibinfo{person}{Mustafa~Feyyaz Sonbudak}, \bibinfo{person}{Giovanni~M Troiano}, {and} \bibinfo{person}{Casper Harteveld}.} \bibinfo{year}{2021}\natexlab{}.
\newblock \showarticletitle{“Why are you playing games? You are a girl!”: Exploring gender biases in Esports}. In \bibinfo{booktitle}{\emph{Proceedings of the 2021 CHI conference on human factors in computing systems}}. \bibinfo{pages}{1--15}.
\newblock


\bibitem[Melhart et~al\mbox{.}(2022)]%
        {melhart2022arousal}
\bibfield{author}{\bibinfo{person}{David Melhart}, \bibinfo{person}{Antonios Liapis}, {and} \bibinfo{person}{Georgios~N Yannakakis}.} \bibinfo{year}{2022}\natexlab{}.
\newblock \showarticletitle{The arousal video game annotation (AGAIN) dataset}.
\newblock \bibinfo{journal}{\emph{IEEE Transactions on Affective Computing}} \bibinfo{volume}{13}, \bibinfo{number}{4} (\bibinfo{year}{2022}), \bibinfo{pages}{2171--2184}.
\newblock


\bibitem[Michele(2023)]%
        {michele_howlongtobeat_2023}
\bibfield{author}{\bibinfo{person}{Michele}.} \bibinfo{year}{2023}\natexlab{}.
\newblock \bibinfo{title}{{HowLongToBeat} {Python} {API}}.
\newblock
\newblock
\urldef\tempurl%
\url{https://github.com/ScrappyCocco/HowLongToBeat-PythonAPI}
\showURL{%
\tempurl}
\newblock
\shownote{original-date: 2018-12-28T22:50:59Z}.


\bibitem[Nardone et~al\mbox{.}(2023)]%
        {10.1145/3563214}
\bibfield{author}{\bibinfo{person}{Vittoria Nardone}, \bibinfo{person}{Biruk Muse}, \bibinfo{person}{Mouna Abidi}, \bibinfo{person}{Foutse Khomh}, {and} \bibinfo{person}{Massimiliano Di~Penta}.} \bibinfo{year}{2023}\natexlab{}.
\newblock \showarticletitle{Video Game Bad Smells: What They Are and How Developers Perceive Them}.
\newblock \bibinfo{journal}{\emph{ACM Trans. Softw. Eng. Methodol.}} \bibinfo{volume}{32}, \bibinfo{number}{4}, Article \bibinfo{articleno}{88} (\bibinfo{date}{may} \bibinfo{year}{2023}), \bibinfo{numpages}{35}~pages.
\newblock
\showISSN{1049-331X}
\urldef\tempurl%
\url{https://doi.org/10.1145/3563214}
\showDOI{\tempurl}


\bibitem[Navarro(2001)]%
        {levenshtein}
\bibfield{author}{\bibinfo{person}{Gonzalo Navarro}.} \bibinfo{year}{2001}\natexlab{}.
\newblock \showarticletitle{A guided tour to approximate string matching}.
\newblock \bibinfo{journal}{\emph{Comput. Surveys}} \bibinfo{volume}{33}, \bibinfo{number}{1} (\bibinfo{year}{2001}), \bibinfo{pages}{31--88}.
\newblock
\urldef\tempurl%
\url{https://doi.org/10.1145/375360.375365}
\showDOI{\tempurl}


\bibitem[Pascarella et~al\mbox{.}(2018)]%
        {pascarella_how_2018}
\bibfield{author}{\bibinfo{person}{Luca Pascarella}, \bibinfo{person}{Fabio Palomba}, \bibinfo{person}{Massimiliano Di~Penta}, {and} \bibinfo{person}{Alberto Bacchelli}.} \bibinfo{year}{2018}\natexlab{}.
\newblock \showarticletitle{How {Is} {Video} {Game} {Development} {Different} from {Software} {Development} in {Open} {Source}?}. In \bibinfo{booktitle}{\emph{2018 {IEEE}/{ACM} 15th {International} {Conference} on {Mining} {Software} {Repositories} ({MSR})}}. \bibinfo{pages}{392--402}.
\newblock
\newblock
\shownote{ISSN: 2574-3864}.


\bibitem[Pathak et~al\mbox{.}(2017)]%
        {pathak_generating_2017}
\bibfield{author}{\bibinfo{person}{Apurva Pathak}, \bibinfo{person}{Kshitiz Gupta}, {and} \bibinfo{person}{Julian McAuley}.} \bibinfo{year}{2017}\natexlab{}.
\newblock \showarticletitle{Generating and {Personalizing} {Bundle} {Recommendations} on {Steam}}. In \bibinfo{booktitle}{\emph{Proceedings of the 40th {International} {ACM} {SIGIR} {Conference} on {Research} and {Development} in {Information} {Retrieval}}} \emph{(\bibinfo{series}{{SIGIR} '17})}. \bibinfo{publisher}{Association for Computing Machinery}, \bibinfo{address}{New York, NY, USA}, \bibinfo{pages}{1073--1076}.
\newblock
\showISBNx{978-1-4503-5022-8}
\urldef\tempurl%
\url{https://doi.org/10.1145/3077136.3080724}
\showDOI{\tempurl}


\bibitem[PlayMyData(2023)]%
        {riccardorubeimsr2024-data-showcase}
\bibfield{author}{\bibinfo{person}{PlayMyData}.} \bibinfo{year}{2023}\natexlab{}.
\newblock \bibinfo{title}{{riccardoRubei}/{MSR2024}-{Data}-{Showcase}: {Repository} for {MSR2024} {Data} {Showcase}}.
\newblock
\newblock
\urldef\tempurl%
\url{https://github.com/riccardoRubei/MSR2024-Data-Showcase}
\showURL{%
\tempurl}


\bibitem[PlaystationStore(2023)]%
        {playstore}
\bibfield{author}{\bibinfo{person}{PlaystationStore}.} \bibinfo{year}{2023}\natexlab{}.
\newblock \bibinfo{title}{Latest {\textbar} {Official} {PlayStation}™{Store}}.
\newblock
\newblock
\urldef\tempurl%
\url{https://store.playstation.com/}
\showURL{%
\tempurl}
\newblock
\shownote{[last accessed on 2023-12-06]}.


\bibitem[Politowski et~al\mbox{.}(2020)]%
        {politowski2020dataset}
\bibfield{author}{\bibinfo{person}{Cristiano Politowski}, \bibinfo{person}{Fabio Petrillo}, \bibinfo{person}{Gabriel~Cavalheiro Ullmann}, \bibinfo{person}{Josias de Andrade~Werly}, {and} \bibinfo{person}{Yann-Ga{\"e}l Gu{\'e}h{\'e}neuc}.} \bibinfo{year}{2020}\natexlab{}.
\newblock \showarticletitle{Dataset of video game development problems}. In \bibinfo{booktitle}{\emph{Proceedings of the 17th International Conference on Mining Software Repositories}}. \bibinfo{pages}{553--557}.
\newblock


\bibitem[Quwaider et~al\mbox{.}(2019)]%
        {quwaider2019impact}
\bibfield{author}{\bibinfo{person}{Muhannad Quwaider}, \bibinfo{person}{Abdullah Alabed}, {and} \bibinfo{person}{Rehab Duwairi}.} \bibinfo{year}{2019}\natexlab{}.
\newblock \showarticletitle{The impact of video games on the players behaviors: A survey}.
\newblock \bibinfo{journal}{\emph{Procedia Computer Science}}  \bibinfo{volume}{151} (\bibinfo{year}{2019}), \bibinfo{pages}{575--582}.
\newblock


\bibitem[Roitberg et~al\mbox{.}(2021)]%
        {roitberg2021let}
\bibfield{author}{\bibinfo{person}{Alina Roitberg}, \bibinfo{person}{David Schneider}, \bibinfo{person}{Aulia Djamal}, \bibinfo{person}{Constantin Seibold}, \bibinfo{person}{Simon Rei{\ss}}, {and} \bibinfo{person}{Rainer Stiefelhagen}.} \bibinfo{year}{2021}\natexlab{}.
\newblock \showarticletitle{Let’s play for action: Recognizing activities of daily living by learning from life simulation video games}. In \bibinfo{booktitle}{\emph{2021 IEEE/RSJ International Conference on Intelligent Robots and Systems (IROS)}}. IEEE, \bibinfo{pages}{8563--8569}.
\newblock


\bibitem[Rubei and Di~Sipio(2021)]%
        {rubei2021auryga}
\bibfield{author}{\bibinfo{person}{Riccardo Rubei} {and} \bibinfo{person}{Claudio Di~Sipio}.} \bibinfo{year}{2021}\natexlab{}.
\newblock \showarticletitle{AURYGA: A Recommender System for Game Tagging.}. In \bibinfo{booktitle}{\emph{IIR}}.
\newblock


\bibitem[Simonyan and Zisserman(2015)]%
        {vggnet}
\bibfield{author}{\bibinfo{person}{Karen Simonyan} {and} \bibinfo{person}{Andrew Zisserman}.} \bibinfo{year}{2015}\natexlab{}.
\newblock \bibinfo{title}{Very Deep Convolutional Networks for Large-Scale Image Recognition}.
\newblock
\newblock
\showeprint[arxiv]{1409.1556}~[cs.CV]


\bibitem[Steam(2023)]%
        {steam}
\bibfield{author}{\bibinfo{person}{Steam}.} \bibinfo{year}{2023}\natexlab{}.
\newblock \bibinfo{title}{Steam {Store}}.
\newblock
\newblock
\urldef\tempurl%
\url{https://store.steampowered.com/}
\showURL{%
\tempurl}
\newblock
\shownote{[last accessed on 2023-12-06]}.


\bibitem[Thompson et~al\mbox{.}(2017)]%
        {thompson2017sentiment}
\bibfield{author}{\bibinfo{person}{Joseph~J Thompson}, \bibinfo{person}{Betty~HM Leung}, \bibinfo{person}{Mark~R Blair}, {and} \bibinfo{person}{Maite Taboada}.} \bibinfo{year}{2017}\natexlab{}.
\newblock \showarticletitle{Sentiment analysis of player chat messaging in the video game StarCraft 2: Extending a lexicon-based model}.
\newblock \bibinfo{journal}{\emph{Knowledge-Based Systems}}  \bibinfo{volume}{137} (\bibinfo{year}{2017}), \bibinfo{pages}{149--162}.
\newblock


\bibitem[Wood(2008)]%
        {wood2008problems}
\bibfield{author}{\bibinfo{person}{Richard~TA Wood}.} \bibinfo{year}{2008}\natexlab{}.
\newblock \showarticletitle{Problems with the concept of video game “addiction”: Some case study examples}.
\newblock \bibinfo{journal}{\emph{International journal of mental health and addiction}}  \bibinfo{volume}{6} (\bibinfo{year}{2008}), \bibinfo{pages}{169--178}.
\newblock


\bibitem[Zadtootaghaj et~al\mbox{.}(2018)]%
        {8463434}
\bibfield{author}{\bibinfo{person}{Saman Zadtootaghaj}, \bibinfo{person}{Steven Schmidt}, \bibinfo{person}{Nabajeet Barman}, \bibinfo{person}{Sebastian Möller}, {and} \bibinfo{person}{Maria~G. Martini}.} \bibinfo{year}{2018}\natexlab{}.
\newblock \showarticletitle{A Classification of Video Games based on Game Characteristics linked to Video Coding Complexity}. In \bibinfo{booktitle}{\emph{2018 16th Annual Workshop on Network and Systems Support for Games (NetGames)}}. \bibinfo{pages}{1--6}.
\newblock
\urldef\tempurl%
\url{https://doi.org/10.1109/NetGames.2018.8463434}
\showDOI{\tempurl}


\end{thebibliography}


\end{document}